\documentclass[11pt]{article}

\usepackage{amssymb}
\usepackage{amsmath}
\usepackage{graphicx} 
\usepackage{tikz}
\usepackage[margin=1in]{geometry}
\usepackage{adjustbox}
\usepackage{float}
\usepackage{booktabs}
\usepackage{authblk}

\title{Rethinking Gaussian-Windowed Wavelets for Damping Identification}

\author[1]{Hadi M. Daniali}
\author[1]{Martin v. Mohrenschildt}
\affil[1]{Department of Computing and Software, McMaster University, 1280 Main St W, Hamilton, ON L8S 4K1, Canada}
\date{}
\begin{document}
\maketitle

\begin{abstract}
In modal analysis, the prevalent use of Gaussian-based wavelets (such as Morlet and Gabor) for damping estimation is rarely questioned. In this study, we challenge this conventional approach by systematically exploring envelope-based damping estimators and proposing a data-driven framework that optimizes the shape and parameters of the envelope utilizing synthetic impulse responses with known ground-truth envelopes. The performance of the resulting estimators is benchmarked across a range of scenarios and compared against frequency-domain damping estimation methods, including Least Squares Rational Function (LSRF), poly-reference Least Squares Complex Frequency-Domain (pLSCF), peak picking (PP), and the Yoshida method.
Our findings indicate that Triangle and Welch windows consistently outperform or are on par with Gaussian wavelet methods in contexts of moderate to high signal-to-noise ratios (SNR). In contrast, Blackman filtering demonstrates superior robustness under low SNR conditions and scenarios involving closely spaced modes. Among the frequency-domain methods assessed, LSRF shows the most reliability at very low SNR; however, the non-Gaussian optimized envelope estimators perform exceptionally well as the SNR improves. 
\end{abstract}

\section{Introduction}

Modal analysis is a technique in engineering that focuses on studying the dynamic properties of mechanical structures under vibration \cite{ewins2009modal}. This method is widely utilized across various disciplines, including civil \cite{chopra2007dynamics}, mechanical \cite{rao2001mechanical}, aerospace \cite{kerschen2013nonlinear}, bioengineering \cite{app132011566}, and automotive \cite{inman1994engineering} engineering, to analyze diverse structures such as bridges, buildings, aircraft, vehicles, and machinery. The primary objective of modal analysis is to identify key parameters, namely natural frequencies, mode shapes, and damping ratios, which are essential for understanding the structural responses to different forces, particularly vibrational loads. This knowledge is imperative for ensuring optimal performance and preventing issues related to resonance and structural failure \cite{HE20011}. Consequently, modal analysis serves a vital role in structural health monitoring, vibration control, performance enhancement, and predictive maintenance \cite{liu2021_ml_modal,santos2020automatic,app10010048,soria2012operational,FENG2013165,zhang2009rolling}.

Two prevalent methodologies for conducting modal analysis are Experimental Modal Analysis (EMA) and Operational Modal Analysis (OMA). EMA consists of artificially exciting a structure with known forces, employing tools such as impact hammers or shakers, and subsequently measuring the structural response to extract the modal parameters \cite{ewins2009modal}. Conversely, OMA, often referred to as ambient modal analysis, capitalizes on the natural operational conditions that structures encounter, such as wind, traffic, or inherent operational vibrations, to measure responses without directly applying external forces \cite{brincker2001modal}. While EMA typically provides greater accuracy in modal parameter identification, OMA is particularly useful for large-scale structures, such as bridges and high-rise buildings, where the application of artificial excitations may be impractical.

Damping refers to the process by which vibrational energy within a structure is dissipated, typically through mechanisms such as friction, material deformation, or other forms of resistance \cite{slavivc2003damping,staszewski1997identification}. In modal analysis, accurately identifying the damping ratio is essential, as it helps predict how quickly a structure will return to its rest state after experiencing a disturbance. Damping plays a critical role in preventing sustained vibrations, which can lead to resonance and potential structural failure.
By continuously monitoring damping ratios, engineers can detect subtle changes in a structure’s dynamic behavior that may indicate damage, degradation, or other anomalies, allowing them to forecast when and where maintenance actions are needed. Modal analysis is considered the most reliable method for extracting damping information \cite{zahid2020review}. However, for even simple linear time-invariant (LTI) systems, this identification process can be more challenging than identifying other modal parameters, such as natural frequencies and mode shapes \cite{daniali2025wavelet}. This difficulty becomes even more pronounced in multi-degree-of-freedom (MDOF) systems and a noisy environment \cite{staszewski1997identification,nadkarni2021experimental,qu2024review}.

Numerous methodologies exist for extracting damping in EMA based on a system's impulse response. These methodologies can be classified into three primary categories: time-domain, frequency-domain, and time-frequency domain techniques. A notable limitation in the existing comparative literature, as highlighted in studies such as \cite{Zrayka2019comparison} and \cite{zielinski2011frequency}, is the assumption of perfect signal alignment. Many investigations presuppose that the impact excitation occurs precisely at the time of the recorded signal, facilitating a more straightforward identification of modal parameters. While this assumption simplifies the analytical process, it is rarely attainable in practical scenarios—particularly when sensors are not directly placed on the device applying the impact.

Additionally, many of these methodologies exhibit considerable performance degradation in the presence of noise. Among the various approaches, wavelet-based techniques have demonstrated notable resilience against noise, especially in lightly damped systems \cite{zielinski2011frequency,boltevzar2004enhancements,tomac2022damping}. In our previous research, we conducted a comprehensive analysis of several well-established damping extraction methods, focusing on multiple non-aligned and noisy impulse responses \cite{daniali2025wavelet}. The findings from this study underscore the efficacy of wavelet-based methods, which operate in the time-frequency domain, despite the challenges posed by noise and non-alignment.

The motivation for employing wavelet-based methods arises from their inherent capability to extract modal envelopes via a linear functional transformation. In particular, the task of identifying damping characteristics from impulse responses requires the extraction of the envelope of each mode individually. To facilitate this process, the recorded MDOF signal must first undergo a linear transformation to yield single-degree-of-freedom (SDOF) modal responses. Following this transformation, the envelope of each mode is derived through an additional linear transformation. As a result, the entire procedure from the MDOF signals to the estimation of modal envelopes is linear and can thus be articulated as a linear functional transform. This is precisely the function that wavelets, along with general envelope estimation techniques, are designed to perform.

Furthermore, the linearity inherent in the entire envelope estimation process indicates that any additive white Gaussian noise (AWGN) present in the original recorded signals will propagate linearly to the extracted envelopes, thereby maintaining the Gaussian characteristics of the noise. Under typical experimental conditions, impulse response measurements are frequently repeated, resulting in multiple signals that encapsulate the same underlying modal response, infected by independent and identically distributed (IID) Gaussian noise. Given the Gaussian distribution and the independence of the noise, averaging the extracted modal envelopes across these repeated recordings effectively diminishes the variance associated with the noise component. Thus, as demonstrated in our previous research \cite{daniali2025wavelet}, the utilization of multiple impulse response measurements combined with envelope averaging presents a straightforward yet robust approach to significantly enhancing both the accuracy and reliability of damping ratio estimates in practical modal analysis applications.

In recent decades, numerous studies have been conducted to enhance the estimation of damping parameters through the application of wavelet transforms \cite{boltevzar2004enhancements,tomac2022damping,lee2021analysis}. Among the various wavelet functions, Morlet and Gabor wavelets, which utilize Gaussian windows, have gained prominence for their optimal balance between temporal and spectral resolution \cite{staszewski1997identification}.
In their comprehensive investigation, Le and Argoul emphasized that the selection of the mother wavelet plays a crucial role in the accuracy of extracting modal parameters, including natural frequencies and damping ratios \cite{le2004continuous}. Their findings highlighted the importance of adjusting the ratio of the center frequency to the frequency bandwidth in this context. While both Morlet and Cauchy wavelets exhibited commendable performance, the Morlet wavelet was found to possess superior time-frequency localization, thereby making it more effective for precise modal analysis.

In their 2018 study, Gaviria and Montejo investigated the application of optimal wavelet parameters for system identification in civil engineering structures, with a particular focus on the continuous time wavelet transform (CTWT) \cite{gaviria2018optimal}. They conducted a comparative analysis of Gabor and complex Morlet wavelets, concluding that both exhibit proportional behavior in terms of coefficients, thereby indicating no significant advantage of one over the other. Their findings emphasize the necessity of selecting optimal wavelet parameters based on the specific identification objective: maximizing time resolution is preferable for identifying instantaneous frequencies, while enhancing frequency resolution proves more effective for estimating average damping ratios. This observation raises an important consideration: although the Morlet wavelet is recognized for its optimal balance of time-frequency localization, the Gaussian window may not universally provide the best conditions for damping extraction. In a related context, Silik et al. demonstrated that despite the time-frequency localization properties being nearly identical for Gaussian window wavelets, certain non-Gaussian-based wavelets exhibited substantial robustness in de-noising, ultimately facilitating clearer interpretations of dynamic structural measurement results \cite{silik2021analytic}.

The objective of this research is to numerically optimize the parameters of the envelope estimator window/filter, thereby enhancing the accuracy of damping extraction in LTI systems, particularly within noisy environments. This study employs numerical methods to fine-tune the parameters of the window/filter, aiming to improve the efficacy of the envelope estimator and mitigate errors associated with damping extraction. The optimization process will involve a comprehensive exploration of various window functions alongside a systematic refinement of their respective parameters.

In a data-driven optimization context, a primary challenge arises from the unavailability of precise modal envelopes and damping parameters. In practical experimental settings, these true parameters are typically unknown, complicating performance evaluation and hindering direct optimization efforts. 
However, leveraging accurate mathematical modeling capabilities for LTI systems allows for the generation of synthetic datasets based on these established models. One significant advantage of utilizing synthetic data is the explicit knowledge of the exact modal parameters and associated labels, including damping ratios and modal envelopes. This understanding facilitates a comprehensive evaluation of the performance of envelope estimators, thereby enabling effective numerical optimization processes.

Synthetic data has gained significant traction across various fields, particularly in the realm of machine learning \cite{nikolenko2021synthetic}. It plays a crucial role in applications such as computer vision \cite{olson2018synthetic} and neural programming \cite{kurach2015neural}, particularly when the availability of real-world data is limited or when the cost of generating labeled datasets is high. Moreover, synthetic data is also extensively utilized in simulated environments such as autonomous driving \cite{lategahn2011visual} and robotics \cite{chociej2019orrb}, where training and tuning control algorithms, like reinforcement learning, are costly and challenging to implement practically.

The remainder of this paper is organized as follows. Section \ref{matmodel} introduces the LTI mechanical modeling framework used throughout the paper, derives the modal impulse-response representation, and summarizes key practical measurement issues that affect damping estimation. Section \ref{methods_sec} reviews representative damping-identification approaches in the time domain, frequency domain, and time–frequency domain, including envelope-based methods. Section \ref{sec:envelope_estimation} presents the proposed envelope-estimation framework and the data-driven optimization procedure used to select the window/filter form and its parameters for accurate damping extraction. Section \ref{sec:numerical_case} reports numerical case studies across three representative scenarios and discusses the estimator’s performance trends under varying conditions. Finally, Section \ref{conclusion} concludes with the main findings and practical recommendations.

\section{Mathematical Background} \label{matmodel}
Before delving deeper into modal analysis techniques, it is essential to establish a foundation by explaining and modeling dynamic structures. The dynamic behavior of structures can often be effectively represented using LTI mechanical models, where the governing equations are based on classical mechanics and linear differential equations. 

\subsection{LTI Mechanical Models}
A standard representation of structural dynamics in modal analysis considers LTI mechanical systems with viscous damping. Such systems are mathematically expressed using second-order differential equations in matrix form, as demonstrated in the literature~\cite{staszewski1997identification,silva2021online}.

\begin{equation}
\mathbf{M}\ddot{\mathbf{x}}(t) + \mathbf{C}\dot{\mathbf{x}}(t) + \mathbf{K}\mathbf{x}(t) = \mathbf{f}(t),
\label{eq:motion}
\end{equation}
where \(\mathbf{M}\) denotes the mass matrix, which encapsulates the inertia properties of the system; \(\mathbf{C}\) represents the damping matrix, reflecting the energy dissipation characteristics; and \(\mathbf{K}\) signifies the stiffness matrix, which characterizes the elastic properties of the structure. The vectors \(\mathbf{x}(t)\), \(\dot{\mathbf{x}}(t)\), and \(\ddot{\mathbf{x}}(t)\) correspond to the displacement, velocity, and acceleration, respectively. Additionally, \(\mathbf{f}(t)\) represents the external force vector that is applied to the structure.

Equation~\eqref{eq:motion} is a coupled set of \(N\) second-order differential equations, which can be computationally demanding to analyze directly. A common simplification is proportional (Rayleigh) damping, which assumes a linear combination of the mass and stiffness matrices~\cite{staszewski1997identification,silva2021online,liu1995formulation}.

\begin{equation}
\mathbf{C} = \alpha\mathbf{M} + \beta\mathbf{K},
\label{eq:rayleigh}
\end{equation}
where \(\alpha\) and \(\beta\) are the Rayleigh damping coefficients that weight the mass-proportional and stiffness-proportional components of \(\mathbf{C}\), respectively.
Substituting Equation~\eqref{eq:rayleigh} into Equation~\eqref{eq:motion} yields:

\begin{equation}
\mathbf{M}\ddot{\mathbf{x}}(t) + (\alpha\mathbf{M} + \beta\mathbf{K})\dot{\mathbf{x}}(t) + \mathbf{K}\mathbf{x}(t) = \mathbf{f}(t).
\label{eq:motion_rayleigh}
\end{equation}

To further simplify analysis, modal decomposition is employed. This involves representing the displacement vector \(\mathbf{x}(t)\) as a linear combination of the mode shapes of the undamped system:

\begin{equation}
\mathbf{x}(t) = \mathbf{\Phi}\mathbf{q}(t),
\label{eq:modal_transformation}
\end{equation}
where \(\mathbf{\Phi}\) is the modal matrix, whose columns are mode shapes (eigenvectors), and \(\mathbf{q}(t)\) is the modal coordinate vector. Substituting Equation~\eqref{eq:modal_transformation} into Equation~\eqref{eq:motion_rayleigh}, and using the orthogonality properties of the mode shapes with respect to the mass and stiffness matrices:

\begin{equation}
\mathbf{\Phi}^{T}\mathbf{M}\mathbf{\Phi} = \mathbf{I}, \quad \mathbf{\Phi}^{T}\mathbf{K}\mathbf{\Phi} = \mathbf{\Omega}^2.
\label{eq:orthogonality_conditions}
\end{equation}
Here, \(\mathbf{I}\) is the identity matrix, and \(\mathbf{\Omega}^2\) is the diagonal matrix containing squared natural frequencies. Therefore, the equation of motion reduces to \(N\) decoupled second-order equations expressed in modal coordinates as follows:

\begin{equation}
\ddot{\mathbf{q}}(t) + (\alpha\mathbf{I} + \beta\mathbf{\Omega}^2)\dot{\mathbf{q}}(t) + \mathbf{\Omega}^2\mathbf{q}(t) = \mathbf{\Phi}^{T}\mathbf{f}(t).
\label{eq:decoupled_modal_eqs}
\end{equation}
Thus, under the proportional damping assumption, the coupled system reduces to \(N\) independent SDOF modal equations that can be solved separately.

\subsection{Impulse Response of Modal Equations}

To understand how a structure responds to excitation, EMA commonly examines impulse responses. Consider an impulse excitation characterized by \(\mathbf{f}(t) = \delta(t)\), where \(\delta(t)\) denotes the Dirac delta function. By defining the modal forces as \(\mathbf{F}_m = \mathbf{\Phi}^{T}\),  each modal equation can be individually expressed as:

\begin{equation}
\ddot{q}_i(t) + 2\zeta_i{\omega_n}_i\dot{q}_i(t) + {\omega_n}_i^2 q_i(t) = F_{m_i}\delta(t), \quad i=1,2,...,N,
\label{eq:single_modal_equation}
\end{equation}
where \({\omega_n}_i\) is the \(i\)-th mode natural frequency, and \(\zeta_i\) is the \(i\)-th mode damping ratio, defined from the Rayleigh damping equation (Equation~\ref{eq:rayleigh}):

\begin{equation}
    2\zeta_i{\omega_n}_i = \alpha + \beta{\omega_n}_i^2.
\end{equation}
For the common case of underdamped structures, where \(0<\zeta_i<1\), the analytical solution to Equation~\eqref{eq:single_modal_equation}, representing the impulse response for each mode, is expressed as follows:

\begin{equation}
q_i(t) = A_i e^{-\zeta_i{\omega_n}_i t}\sin({\omega_d}_i t), \quad t\geq 0,
\label{eq:impulse_response_modal}
\end{equation}
where the modal amplitude \( A_i \) is given by:

\begin{equation}
    A_i = \frac{F_{m_i}}{{\omega_d}_i},
\end{equation}
and \({\omega_d}_i = {\omega_n}_i\sqrt{1 - \zeta_i^2}\) is the damped natural frequency of mode \(i\).

Using Equation~\eqref{eq:modal_transformation}, the total impulse response in physical coordinates \(\mathbf{x}(t)\) becomes the superposition of all modal responses:

\begin{equation}
    \mathbf{x}(t) = \sum_{i=1}^{N}\mathbf{\phi}_i q_i(t) = \sum_{i=1}^{N}\mathbf{\phi}_i A_i e^{-\zeta_i{\omega_n}_i t}\sin({\omega_d}_i t), \quad t\geq 0.
    \label{eq:impulse_response_total}
\end{equation}
Thus, with the proportional damping assumption, the initially coupled system is simplified into \(N\) analytically solvable and independent equations, facilitating deeper analysis and understanding of structural dynamics.

\subsection{Observation Model and Practical Challenges}\label{modelobs}

In real-world experimental conditions, damping identification is affected by numerous sources of uncertainty and deviations from theoretical assumptions. Typically, modal analysis relies on multiple recordings obtained from repeated impacts, multiple sensors, or both. While these multiple measurements enhance robustness and statistical reliability, they introduce additional practical challenges. The measured responses often deviate significantly from ideal theoretical models due to instrumentation limitations, signal distortions, and structural complexities. The following challenges are commonly encountered across damping estimation methods:

\begin{itemize}
    \item \textbf{Unknown excitation amplitude}:
    When the input force is applied using devices lacking calibrated force sensors, the true excitation amplitude remains unknown. This introduces an unknown scaling factor into the measured response, complicating quantitative interpretation and comparison.

    \item \textbf{Unknown impact time (time shift)}:
    The precise timing of the excitation is often not known, especially when triggers are asynchronous or the recording begins before the impact occurs. This results in an unknown time shift (\(\tau_0\)) in the recorded signal, which may misalign the response relative to time-domain models.
    
    \item \textbf{Measurement noise}:
    Noise from sensors, environmental conditions, or electronic interference contaminates the signal and reduces the reliability of modal parameter estimates. This is particularly problematic at low signal-to-noise ratio (SNR) levels, where noise can obscure modal decay characteristics.
    
    \item \textbf{Edge effects} \textit{(specific to envelope-based methods)}:
    In envelope-based techniques such as wavelet analysis, edge effects arise due to the finite length of the signal. Near the boundaries, these methods distort the shape of the estimated decay envelope and can bias damping estimates if not handled appropriately.

   \item \textbf{Presence of closely spaced modes}:
    When modes are very close in frequency, their contributions to the response may overlap, making it difficult to isolate and estimate damping for a single mode. This challenge affects both time- and frequency-domain approaches and requires high (spectral) selectivity by the method to distinguish between modes.

    \item \textbf{Spectral leakage}:
    Because real signals are finite in length, they are not perfectly periodic within the observation window. This results in spectral leakage, where modal energy spreads into adjacent frequencies. Leakage is particularly problematic when estimating closely spaced modes or using frequency-domain methods.
    
    \item \textbf{Multiple measurements}:
    In test scenarios involving multiple impacts or multiple sensors on the same system, slight differences in trigger timing or acquisition delays can cause misalignment between recordings. This can introduce inconsistencies and reduce estimation accuracy, especially when statistical or ensemble-based techniques are used.
    
    \item \textbf{Numerical resolution limitations}:
    The sampling rate and the number of samples determine the time and frequency resolution of the analysis. Insufficient resolution may result in under-resolved modal decays or coarse spectral features, thereby degrading the accuracy of damping identification.

\end{itemize}
These challenges are inherent to EMA and must be taken into account when designing tests, preprocessing signals, and selecting estimation methods.

Thus, each measured response (denoted \(\mathbf{Signal}(t)\)) can be more realistically modeled as:

\begin{equation}
    \mathbf{Signal}(t) = B\,\mathbf{x}(t - \tau_0) + \boldsymbol{\epsilon}(t), \quad t \geq 0,
    \label{eq:observation_model}
\end{equation}

where:
\begin{itemize}
    \item \(B\) represents an unknown scaling factor arising from the unknown excitation amplitude.
    \item \(\tau_0\) is an unknown time shift reflecting the uncertainty in the exact timing of the impact.
    \item \(\boldsymbol{\epsilon}(t)\) denotes additive measurement noise, typically modeled as a zero-mean (often Gaussian) random process.
\end{itemize}

When extracting modal parameters from experimental data, it is important to consider these practical factors, as they can introduce errors into the estimation process.

\section{Damping Identification Methods} \label{methods_sec}

Damping identification is a fundamental task in structural dynamics and modal analysis, providing critical insights into the energy dissipation characteristics of mechanical systems. Accurate estimation of damping ratios improves prediction of system behavior under dynamic loading, informs design modifications for vibration control, and supports condition monitoring and fault detection. Various approaches have been developed to identify damping parameters from experimental data, broadly categorized into time-domain, frequency-domain, and time-frequency domain techniques. Each class offers unique advantages and limitations. This section reviews prominent damping identification methodologies and outlines their theoretical foundations.

\subsection{Time-Domain Damping Identification Using Impulse Responses}
\label{sec:timemethods}

Time-domain methods estimate damping directly from measured time responses (e.g., free decays following impulse or ambient excitation). Their complexity and suitability depend on the system characteristics and the quality of the available data.

\textit{Logarithmic Decrement} is a simple yet effective technique widely used to lightly damped systems. It estimates the damping ratios directly from the amplitude decay observed between successive response peaks in free vibration \cite{cooper1996extending}.

\textit{Ibrahim Time Domain (ITD)} constructs a generalized eigenvalue problem by forming and comparing time-shifted response matrices. The solution to this eigenvalue problem yields complex eigenvalues from which damping and frequency parameters are directly determined \cite{qu2024review,brincker2019modal}.

The \textit{Eigensystem Realization Algorithm (ERA)} involves the construction of Hankel matrices from measured impulse responses, followed by singular value decomposition (SVD) to identify system eigenvalues \cite{caicedo2011practical}.

\textit{Empirical Mode Decomposition (EMD)} offers flexibility by decomposing the impulse response into intrinsic mode functions (IMFs). Each IMF approximately represents a specific frequency band, enabling damping estimation using other time-domain methods \cite{rilling2003empirical}.

\textit{Least Squares Complex Exponential (LSCE)} fits the impulse response data as a sum of exponentially decaying sinusoidal signals. By employing a least-squares optimization approach, LSCE determines damping ratios and modal frequencies, making it particularly suitable for MDOF systems\cite{qu2024review, allemang1987experimental, chen2012comparison}.

\textit{AutoRegressive Moving Average (ARMA)} models the measured impulse response as a combination of autoregressive (AR) and moving-average (MA) processes. By solving the AR characteristic polynomial, ARMA directly identifies system poles, enabling estimation of damping ratios and natural frequencies \cite{vu2007multi}.

Finally, the \textit{Random Decrement Technique (RDT)} iteratively refines modal parameters. This recursive updating facilitates the identification of damping characteristics, which is particularly beneficial in nonlinear systems \cite{he2011emd}.

\subsection{Frequency-Domain Damping Identification Methods}

Frequency-domain damping identification methods are essential tools in both OMA and EMA. These techniques utilize frequency response data to extract modal parameters, including natural frequencies, damping ratios, and mode shapes.

\textit{Peak Picking (PP)} assumes that each dominant peak in the frequency response function (FRF) is dominated by a single mode. While basic peak picking estimates damping using the half-power bandwidth method, advanced approaches refine this assumption. 

To explain the half-power bandwidth method mathematically, we begin with the definition of the quality factor ($Q$), which measures how sharply the resonance is. The quality factor is defined as:
\begin{equation}
\label{eq:q}
Q = \frac{f_r}{\Delta f},
\end{equation}
where $f_r$ is the resonance frequency, and $\Delta f$ is the bandwidth at which the amplitude drops to \(\frac{1}{\sqrt{2}}\) of its peak, also known as the half-power points \cite{wang2011analysis}.
The damping ratio ($\zeta$) is inversely related to the quality factor and can be expressed as:
\begin{equation}
\label{eq:zetafromq}
\zeta = \frac{1}{2Q} = \frac{\Delta f}{2f_r}.
\end{equation}

One enhanced version of peak picking fits a local FRF section around each peak to an SDOF damped harmonic oscillator model using a least-squares curve fit. This method solves a system of equations to estimate the resonance frequency and damping ratio, improving accuracy compared with the basic half-power technique \cite{brandt2023noise}.

\textit{pLSCF (poly-reference Least-Squares Complex Frequency-Domain)} (also known as PolyMAX) fits multiple FRFs simultaneously using a rational polynomial model in the complex frequency domain. It estimates both numerator and denominator coefficients of that model, enabling extraction of poles and modal parameters, including damping ratio \cite{peeters2004polymax}.

\textit{Least Squares Rational Function (LSRF)} method is structurally similar to pLSCF but applies nonlinear least-squares fitting to the magnitude of the FRF rather than the complex-valued FRF \cite{ozdemir2017transfer}.

\textit{Bertocco-Yoshida} local curve-fitting method. It uses three frequency points around a resonance peak and solves a closed-form system for the three parameters of a second-order SDOF rational model. This closed-form approach is computationally efficient and can provide accurate damping estimates \cite{duda2013efficacy}.

\subsection{Wavelet-Based Damping Identification} \label{sec:wavelet}

Wavelet analysis is a powerful mathematical tool that enables simultaneous signal analysis in both time and frequency domains. Unlike the traditional Fourier transforms, which provide frequency information with no temporal localization, wavelet transforms offer a localized time-frequency decomposition, making them particularly effective for analyzing transient and non-stationary signals, such as those encountered in impact modal analysis.

The fundamental element of wavelet analysis is the \emph{wavelet}, denoted here by the mother wavelet \(\Psi(t)\). A mother wavelet should satisfy the following essential mathematical properties:

Absolute and square integrability:
    \begin{equation}
        \int_{-\infty}^{\infty} |\Psi(t)|\, dt < \infty, \quad\quad
        \int_{-\infty}^{\infty} |\Psi(t)|^2\, dt < \infty.
    \end{equation}
We further impose zero mean and unit-energy normalization:
    \begin{equation}
    \label{wavelet_condition}
        \int_{-\infty}^{\infty}\Psi(t)\,dt = 0, \quad\quad
        \int_{-\infty}^{\infty}|\Psi(t)|^2\,dt = 1.
    \end{equation}

A family of wavelets is generated by scaling and translating the mother wavelet \(\Psi(t)\):

\begin{equation}
\Psi_{s,\tau}(t) = \frac{1}{\sqrt{s}}\Psi\left(\frac{t - \tau}{s}\right), \quad s,\tau \in \mathbb{R}, \quad s > 0,
\end{equation}
where \(s\) represents the scale factor controlling the dilation (frequency resolution), and \(\tau\) denotes the translation factor controlling the time shift.
The continuous wavelet transform (CWT) of a signal \(x(t)\) with respect to the mother wavelet \(\Psi(t)\) is defined by:

\begin{equation}
\mathcal{W}_{\Psi,x}(s,\tau) = \int_{-\infty}^{\infty} x(t)\frac{1}{\sqrt{s}}\overline{\Psi}\left(\frac{t - \tau}{s}\right) dt,
\label{eq:CWT}
\end{equation}
where \(\overline{\Psi}(t)\) denotes the complex conjugate of \(\Psi(t)\). By evaluating this transform across multiple scales \(s\) and translations \(\tau\), the wavelet transform yields a detailed representation of the time-frequency characteristics of \(x(t)\).

A common choice of mother wavelet in modal analysis is the complex Morlet wavelet, which is widely used due to its excellent time-frequency localization and strong resemblance to structural vibration modes~\cite{staszewski1997identification}. Mathematically, the complex Morlet wavelet is expressed as:
\begin{equation}
\Psi(t) = C_{\omega_o} (\sigma^2\pi)^{-\frac{1}{4}} e^{-\frac{t^2}{2\sigma^2}} \left(e^{i\omega_o t} - K_{\omega_o}\right),
\label{morlet_centered}
\end{equation}
where \(\omega_o\) is the central frequency, \(\sigma\) controls the Gaussian window width, and \(C_{\omega_o}\) and \(K_{\omega_o}\) are normalization constants that ensuring the wavelet has unit energy and zero mean (Equation \eqref{wavelet_condition}), respectively.

A simpler, commonly used alternative is the Gabor wavelet, which similarly uses a Gaussian window modulated by a complex exponential but omits the zero-mean normalization terms. Despite lacking strict adherence to classical wavelet definitions, the Gabor wavelet typically provides comparable performance in damping estimation, as modal parameter estimation primarily relies on relative frequency and damping characteristics rather than absolute wavelet amplitude.

Wavelet-based damping ratio estimation leverages the time-frequency localization properties of wavelets to isolate modal responses and quantify their decay rates. In particular, wavelets act as tunable band-pass filters, enabling the extraction of oscillatory components centered around specific frequencies. This is particularly advantageous in impact modal analysis, where multiple modes may be excited simultaneously, and the ability to decompose overlapping modal responses is essential for accurate damping estimation.

From a computational standpoint, the continuous wavelet transform in Equation \eqref{eq:CWT} can be efficiently implemented in the frequency domain via the convolution theorem. Specifically, convolution in the time domain corresponds to pointwise multiplication in the frequency domain:

\begin{equation}
\mathcal{F}\bigl( x(t) *\overline{\Psi}_{s,\tau}(t) \bigl)(\omega) = \mathcal{F}\bigl(x(t)\bigl)(\omega) \cdot \mathcal{F}\bigl(\overline{\Psi}_{s,\tau}(t)\bigl)(\omega),
\end{equation}
where \(*\) denotes convolution and \(\mathcal{F}\bigl(\cdot\bigl)\) is the Fourier transform operator. Thus, the wavelet transform can be computed as:

\begin{equation}
\mathcal{W}_{\Psi,x}(s,\tau) = \mathcal{F}^{-1} \Bigl( \mathcal{F}\bigl(x(t)\bigl)(\omega) \cdot \mathcal{F}\bigl( \overline{\Psi}_{s,\tau}(t) \bigl)(\omega) \Bigl).
\end{equation}
By applying the convolution theorem, the computational complexity can be reduced to \(O(n \log n)\).

In this framework, the wavelet transform acts as a frequency-selective band-pass filter. The Fourier transform of Gaussian-windowed (e.g., Morlet/Gabor) wavelets is typically concentrated around a central frequency \(\omega_0\), with its bandwidth determined by the wavelet’s time spread parameter \(\sigma\). As a result, multiplication in the frequency domain effectively suppresses all frequency components outside the vicinity of \(\omega_0\). This makes the wavelet transform functionally equivalent to a band-pass filter that removes all modal contributions except for the one near the center frequency \(\omega_0\). In modal analysis, this property is exploited to isolate individual vibrational modes, thereby allowing accurate estimation of mode-specific damping ratios.

Once the desired modal component is isolated, the damping ratio can be estimated from the time-domain envelope of the filtered signal. The complex-valued wavelet coefficients yield instantaneous amplitude and phase information. Taking the natural logarithm of the envelope renders the decay linear in time:

\begin{equation} \label{eq:waveletzeta}
\ln \left( |\mathcal{W}_{\Psi,x}(s,\tau)| \right) \approx -\zeta \omega_n t + C,
\end{equation}

where \(C\) is a constant. Linear regression on this log-transformed envelope yields an estimate of the damping ratio from the slope.

Fundamentally, wavelet methods act as \textbf{envelope estimators}: by filtering the measured signal, they isolate and estimate the amplitude envelope associated with a particular vibration mode, which directly relates to the mode's energy decay. As shown in Equation \eqref{eq:waveletzeta}, the envelope obtained through such filtering reflects how the vibrational energy dissipates over time, enabling subsequent damping ratio estimation.

\subsection{Hilbert-Based and Hilbert–Huang Damping Estimation}

The Hilbert transform is a classical signal processing tool that, like wavelet transforms, enables localized time-frequency analysis. It can be interpreted as a specific type of wavelet transform, where the wavelet is defined in the frequency domain as:

\begin{equation}
\Psi_{\text{Hilbert}}(\omega) = -i \ \mathrm{sgn}(\omega),
\end{equation}
where \(\mathrm{sgn}(\omega)\) is the sign function. This frequency-domain operator induces a 90-degree phase shift to all frequency components of the signal, thereby transforming a real-valued input into its quadrature pair. When the Hilbert transform is applied to a real signal \(x(t)\), it produces the so-called \emph{analytic signal}:

\begin{equation}
z(t) = x(t) + i \ \mathcal{H} \bigl( x(t) \bigl),
\end{equation}
where \(\mathcal{H}\bigl( x(t) \bigl),\) denotes the Hilbert transform of \(x(t)\). The envelope of the signal can then be computed as the magnitude of the analytic signal, \(|z(t)|\), i.e., the instantaneous amplitude. In the case of an SDOF system, this envelope exhibits exponential decay, allowing damping ratio estimation through logarithmic decrement:

\begin{equation}
\ln(|z(t)|) \approx -\zeta \omega_n t + C.
\end{equation}

For MDOF systems with overlapping modes, direct application of the Hilbert transform is generally insufficient, as the analytic signal will represent a superposition of modes and yield inaccurate damping estimates. To address this, the signal can first be decomposed using band-pass filters to isolate individual modal components. When properly designed, this process is essentially equivalent to wavelet-based mode separation, where each filter or window isolates a single modal frequency band prior to applying the Hilbert transform.

An alternative and more adaptive approach for modal decomposition is the \emph{Hilbert–Huang Transform (HHT)}, which combines the Hilbert transform with EMD. As mentioned in Section \ref{sec:timemethods}, EMD is a data-driven, nonlinear technique that decomposes a signal into a finite set of IMFs, each representing a narrowband oscillatory component. Although EMD lacks a rigorous theoretical foundation and does not correspond to a well-defined filter bank, it effectively acts as an adaptive decomposition technique that extracts mode-like components without requiring predefined basis functions.

Once the signal is decomposed into IMFs, the Hilbert transform is applied to each IMF, yielding instantaneous amplitude and instantaneous frequency. The damping ratio for each component can then be estimated by linear regression on the logarithm of the amplitude envelope, similar to the procedure in wavelet and Hilbert transform methods.

\section{LTI Systems Envelope Estimation}\label{sec:envelope_estimation}

As discussed in Section~\ref{sec:wavelet}, wavelet-based damping identification fundamentally involves estimating the time-domain envelope of each mode within impulse responses. Traditional wavelet-based damping estimation methods predominantly utilize Gaussian-windowed wavelets, such as Morlet or Gabor wavelets, due to their superior time-frequency localization properties. However, this choice of wavelet window shape is often heuristic and may not be optimal across all signal types or measurement conditions. For a given mode, an ideal envelope estimator, implemented as a time-domain window or a frequency-domain filter, must effectively suppress non-target modes while preserving the frequency response of the target mode. Preservation is most critical near the mode's peak and half-power bandwidth region, ensuring that the resonance sharpness, quantified by the quality factor \(Q\), remains unchanged. Any distortion introduced by the estimator in these critical regions can bias the estimation of the bandwidth \(\Delta f\) and consequently affect the accuracy of the damping ratio \(\zeta\), as defined previously by Equations~\eqref{eq:q} and~\eqref{eq:zetafromq}.

The primary goal of this research is to accurately extract damping ratios from multiple measured impulse responses of the same LTI system, each of which conforms to the observation model given by Equation~\eqref{eq:observation_model}. To overcome the limitations associated with heuristic choices of wavelet shape, we propose a systematic, data-driven optimization framework for designing improved envelope estimators. Our core approach involves optimizing the shape and parameter \(\theta\) of an envelope estimator function \(\Psi(t;\theta)\), so that filtering a measured vibration signal yields an estimated envelope \(\hat{\mathcal{A}}(t)\) closely matching the true amplitude envelope \(\mathcal{A}(t)=\mathbf{\phi} A e^{-\zeta{\omega_n} t}\) of the target mode.

The estimated envelope is obtained as:
\begin{equation}\label{eq:envelope_estimate}
\hat{\mathcal{A}}(t) = \left| x(t) * \Psi(t; \theta) \right|.
\end{equation}
Since envelope estimators may scale the amplitude, both the estimated and true envelopes are normalized to unity at their initial point. This normalization ensures that the optimization targets the exponential decay, rather than amplitude scaling, thereby preserving the decay rate (and associated damping ratio).

Additionally, due to edge effects, it is neither necessary nor beneficial to utilize the entire extracted envelope for damping estimation. We therefore restrict the loss to a suitable segment of the envelope:

\begin{equation}\label{eq:optimized_loss_interval}
(\Psi_{\text{opt}}, \theta_{\text{opt}}) = \arg\min_{\Psi \in \mathcal{W}, \theta} \frac{1}{t_2 - t_1}\int_{t_1}^{t_2}\left(\hat{\mathcal{A}}(t)-\mathcal{A}(t)\right)^2 dt,
\end{equation}
As shown in Equation~\eqref{eq:waveletzeta}, extracting damping parameters uses the log-envelope. Generally, a longer envelope segment improves reliability in damping estimates. However, the logarithmic operation significantly amplifies noise at low amplitudes, introducing numerical instabilities. Consequently, \(t_2\) should be selected before the envelope amplitude becomes too small, and \(t_1\) chosen to avoid edge transients while ensuring a sufficiently long analysis window.

While some sources suggest using at least five oscillation cycles to estimate damping from the envelope reliably \cite{turunen2010selecting}, the literature is inconsistent. We therefore adopt ten cycles to provide a conservative margin, improving estimation robustness and reducing variance. Accordingly, \(t_2\) is chosen just before the envelope amplitude approaches zero to avoid numerical issues in the logarithm, and \(t_1\) is set to ten oscillation cycles before \(t_2\), ensuring a sufficiently long segment for stable damping estimation.

In practical scenarios, real-world signals are digitally recorded, necessitating the discrete formulation of the optimization criterion. Accordingly, Equation~\eqref{eq:optimized_loss_interval} is discretized as follows:
\begin{equation}\label{eq:optimized_loss_discrete}
(\Psi_{\text{opt}}, \theta_{\text{opt}}) = \arg\min_{\Psi \in \mathcal{W}, \theta} \frac{1}{N_2 - N_1 + 1}\sum_{n=N_1}^{N_2}\left(\hat{\mathcal{A}}[n]-\mathcal{A}[n]\right)^2,
\end{equation}
where \(N_1\) and \(N_2\) are the sample indices corresponding to continuous time points \(t_1\) and \(t_2\), respectively.

When actual impulse responses are measured from a system, the true envelope \(\mathcal{A}(t)\) is unknown, which poses a fundamental challenge to the optimization procedure. To facilitate the optimization and evaluation of envelope estimators, we generate a synthetic dataset using Equation~\eqref{eq:impulse_response_total}. The primary advantage of utilizing synthetic data here is that the ground-truth modal parameters and, critically, the true envelopes of each mode are known exactly.

A critical consideration when working with synthetic data is whether it closely represents real-world data. In the context of evaluating envelope estimator performance on LTI systems, the models provided in Section \ref{matmodel} are well-known and established \cite{rao2001mechanical}. Consequently, the synthetic data generated from these established models accurately reflect the impulse responses of LTI systems, and thus provide a reliable basis for performance evaluation and optimization.

Given real impulse responses, traditional modal analysis techniques can be utilized to accurately estimate the system's damped natural frequencies, denoted as \(\boldsymbol{\omega}_d = [\omega_{d_1}, \omega_{d_2}, \ldots, \omega_{d_N}]\). Since these frequency estimation methods are highly reliable and well-established, we assume that the modal frequencies are known precisely and use them directly as center frequencies for the envelope estimators. Moreover, based on prior knowledge or preliminary measurements, we assume that the damping ratios lie within known bounds, \(\zeta_{\text{min}} \leq \boldsymbol{\zeta} = [\zeta_1, \zeta_2, \dots, \zeta_N] \leq \zeta_{\text{max}}\).

Therefore, each synthetic signal is generated with precisely known modal frequencies and random damping ratios \(\boldsymbol{\zeta}\) sampled uniformly from their prescribed intervals, while modal amplitudes are drawn as
\(
\phi_i A_i \sim \mathcal{U}\!\bigl(\phi_{\min}A_{\min},\,\phi_{\max}A_{\max}\bigr).
\)
The frequencies are held constant across all samples. This design choice is intentional: the parameters of the envelope estimator, \(\theta\) (such as window/filter length and shape), depend strongly on the specific mode being estimated and its spectral separation from neighboring modes to be suppressed.
By contrast, amplitude variations do not materially impact the accuracy of the envelope-based damping estimate. They uniformly scale the envelope without affecting the half-power bandwidth or the exponential decay rate; thus the extracted damping ratio remains unchanged. We empirically verified this consistency in our tests and experiments.

\begin{figure}[htbp]\label{flowchart}
    \centering
    \includegraphics[width=0.9\linewidth]{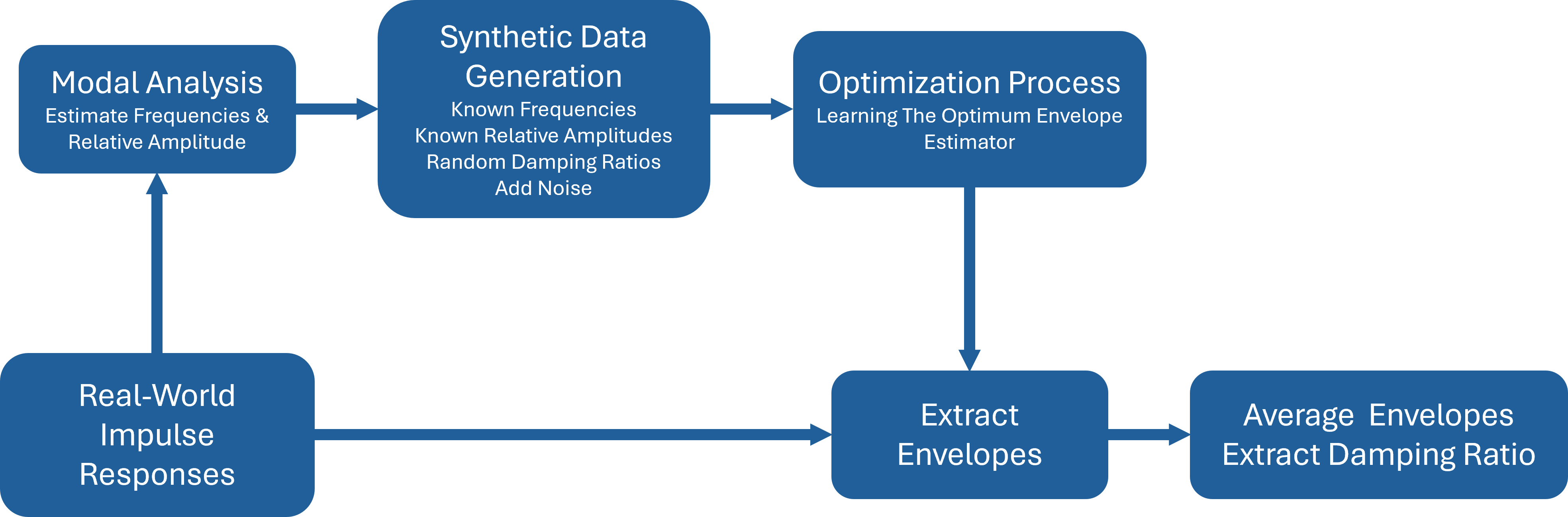}
    \caption{Overview of the optimization process for envelope estimation using synthetic data.}
    \label{fig:optimization_flowchart}
\end{figure}

Additionally, we add AWGN to each synthetic signal to simulate measurement noise, with the SNR (specified in dB unless stated otherwise) randomly sampled from a predefined interval,
\(
\text{SNR} \sim \mathcal{U}(\text{SNR}_{\text{min}}, \text{SNR}_{\text{max}}).
\) Figure~\ref{flowchart} shows an overview of this process. 

Table~\ref{tab:estimators} summarizes the set of window and filter types evaluated in this study.
\begin{table}[H]
\centering
\caption{Envelope estimator types and their formulas (centered at \(t=0\) or \(\omega=0\)).}
\label{tab:estimators}
\renewcommand{\arraystretch}{1.5}
\begin{tabular}{@{}lll@{}}
\toprule
\textbf{Estimator Type} & \textbf{Formula (time or frequency domain)} & $\theta$ \\ 
\midrule
Gaussian Window & \( e^{-\frac{t^2}{2 \sigma^2}} \) & \(\sigma\) \\[6pt]

Rectangular Window & \( \mathrm{rect}\left(\frac{t}{L_t}\right) \) & \(L_t\) \\[6pt]

Shannon Filter & \( \mathrm{rect}\left(\frac{\omega}{L_\omega}\right) \) & \(L_\omega\) \\[6pt]

Triangular Filter & 
\( \left(1 - \left|\frac{2\,\omega}{L_\omega}\right|\right)\cdot\mathrm{rect}\left(\frac{\omega}{L_\omega}\right) \) & \(L_\omega\) \\[6pt]

Triangular Window & 
\( \left(1 - \left|\frac{2\,t}{L_t}\right|\right)\cdot\mathrm{rect}\left(\frac{t}{L_t}\right) \) & \(L_t\) \\[6pt]

Welch Filter & 
\( \left(1 - (\frac{2\,\omega}{L_\omega})^2\right)\cdot\mathrm{rect}\left(\frac{\omega}{L_\omega}\right) \) & \(L_\omega\) \\[6pt]

Welch Window & 
\( \left(1 - (\frac{2\,t}{L_t})^2\right)\cdot\mathrm{rect}\left(\frac{t}{L_t}\right) \) & \(L_t\) \\[6pt]

Blackman Filter\(^*\) & 
\(\left[a_0 - a_1 \cos\left(\frac{2\pi(\omega + L_\omega/2)}{L_\omega}\right) + a_2 \cos\left(\frac{4\pi(\omega + L_\omega/2)}{L_\omega}\right)\right]\cdot\mathrm{rect}\left(\frac{\omega}{L_\omega}\right)\) & \(L_\omega\) \\[6pt]

Blackman Window & 
\(\left[a_0 - a_1 \cos\left(\frac{2\pi(t + L_t/2)}{L_t}\right) + a_2 \cos\left(\frac{4\pi(t + L_t/2)}{L_t}\right)\right]\cdot\mathrm{rect}\left(\frac{t}{L_t}\right)\) & \(L_t\) \\
\bottomrule
\end{tabular}

\vspace{0.8em}
\footnotesize{\(^*\)Blackman coefficients: \(a_0=\frac{7938}{18608}\), \(a_1=\frac{9240}{18608}\), \(a_2=\frac{1430}{18608}\).}

\vspace{0.6em}
\footnotesize{The rectangular function is defined as: 
\(\mathrm{rect}(x)=\begin{cases}1, &|x|\leq 0.5\\0, &|x|>0.5\end{cases}\)}
\end{table}

\section{Numerical Case Study and Discussion}\label{sec:numerical_case}

In this section, we evaluate the envelope estimation framework on three distinct scenarios that vary in different modal frequencies, modal amplitudes, and damping ratios, as summarized in Table~\ref{tab:scenarios}.

\begin{table}[H]
\label{tab:scenarios}
\centering
\caption{Modal parameters for the three test scenarios.}
\renewcommand{\arraystretch}{1.12}
\begin{tabular}{@{}lccc@{}}
\toprule
\textbf{Parameter / Mode} & \textbf{Scenario 1} & \textbf{Scenario 2} & \textbf{Scenario 3} \\
\midrule
\textbf{Frequency (Hz)} & & & \\
\quad Mode 1 & 3.27 & 1.15 & 3.27 \\
\quad Mode 2 & 15.56 & 6.33 & 15.56 \\
\quad Mode 3 & 26.50 & 10.95 & 26.50 \\
\midrule
\textbf{Modal Amplitude } & & & \\
\quad Mode 1 & 1.50 & 2.80 & 1.50 \\
\quad Mode 2 & 2.50 & 1.50 & 2.50 \\
\quad Mode 3 & 1.00 & 4.70 & 1.00 \\
\midrule
\textbf{Damping Ratio (\%)} & & & \\
\quad Mode 1 & 1.50 & 1.50 & 3.50 \\
\quad Mode 2 & 1.00 & 1.00 & 4.00 \\
\quad Mode 3 & 0.80 & 0.80 & 3.00 \\
\bottomrule
\end{tabular}
\end{table}
We designed these three scenarios to span representative conditions encountered in practice, where real systems typically include both lightly and strongly damped modes. Scenario 1 presents well-separated modes with light damping and balanced amplitudes and serves as a baseline. Scenario 2 shifts the spectrum to lower frequencies with more closely spaced modes and reverses amplitude dominance to stress band-pass selectivity and mode separation under modal overlap. Scenario 3 retains the modal frequencies of Scenario 1 but increases damping to test robustness to bandwidth broadening (i.e., a strong-damping regime). Together, these scenarios exercise the estimator across frequency spacing, relative amplitude contrast, and damping level, ensuring that the algorithm is evaluated under both weak- and strong-damping conditions, capturing cases where mode overlap and Q-factor reduction challenge envelope-based methods.

Impulse response signals were generated using the observation model in Equation~\eqref{eq:observation_model}, simulating an LTI system with the modal characteristics specified above; we refer to this as \textbf{Dataset~1}. Each signal is sampled at \(F_s=800\,\mathrm{Hz}\) with a length of \(4096\) point. Additionally, an unknown amplitude  \(B\) and time shift \(\tau_0 \) are drawn independently from
\(B \sim \mathcal{U}(1,5)\) and \(\tau_0 \sim \mathcal{U}(0,2)\,\mathrm{s}\). This sampling frequency rate satisfies the Nyquist criterion and conforms to recommendations to sample at least ten times the highest modal frequency~\cite{d2022minimum}. The primary objective for Dataset~1 is to estimate the damping ratio of the second mode. To assess estimator robustness, we add AWGN with SNR ranging from \(-5\) to \(30\,\mathrm{dB}\).

To optimize the envelope estimator, a larger synthetic dataset of 8,500 impulse responses was generated according to Equation~\eqref{eq:impulse_response_total}, hereafter called \textbf{Dataset~2}. Since the current methods are accurate in estimating the modal frequencies, the true modal frequencies are used to create this dataset. Assuming lightly damping, damping ratios \(\zeta\) were sampled uniformly from the interval \([0.1\%,\, 10\%]\), while modal amplitudes \(\phi A\) were independently sampled from the range \([1,\, 5]\). To simulate realistic measurement conditions, we AWDN with SNR between 10 and 30 dB.

Dataset~2 was split into 7,000 training samples and 1,500 validation samples. The true modal envelopes are analytically available, facilitating direct evaluation of the optimization loss in Equation~\eqref{eq:optimized_loss_interval}. To solve these nonconvex optimization problems, we used gradient descent via backpropagation. We tuned the learning rate, and \(\eta = 0.01\) yielded the best validation performance. Each estimator was trained 15 times with different random initializations to mitigate convergence to local minima, and the best-performing solution for each estimator on the validation set was retained for further analysis.

After determining the optimal \(\theta\) for each window or filter using Dataset~2, these optimized configurations were evaluated on Dataset~1. For each test case, 10-100 independent recordings were available; envelopes were extracted for each recording, temporally aligned at the global maximum sample, and then averaged to reduce noise. The damping ratio was subsequently estimated from the averaged envelope. This ensemble-based averaging procedure follows the methodology in our previous work~\cite{daniali2025wavelet}.

\begin{figure}[H]
  \centering
  \includegraphics[width=\linewidth]{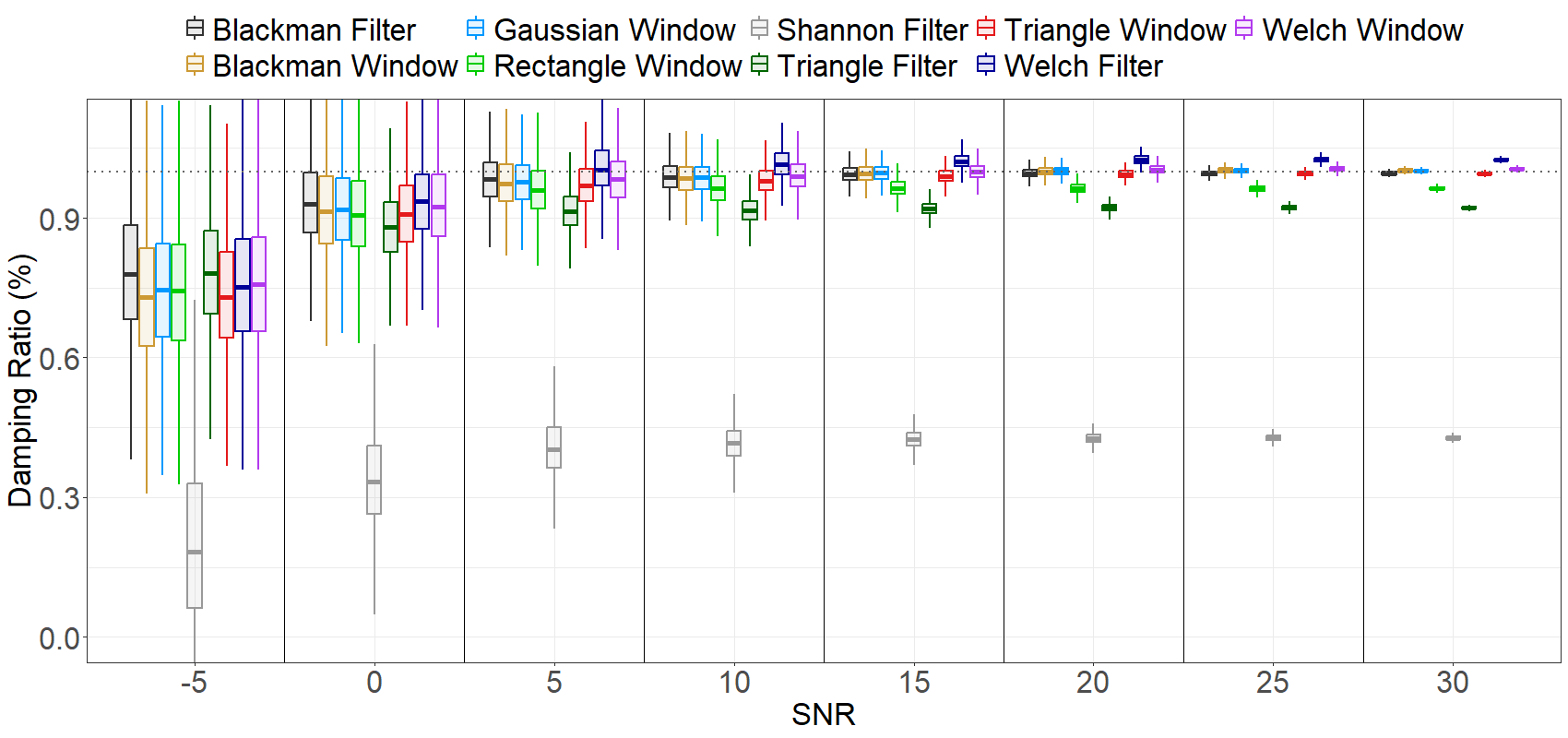}

  \medskip

  \includegraphics[width=\linewidth]{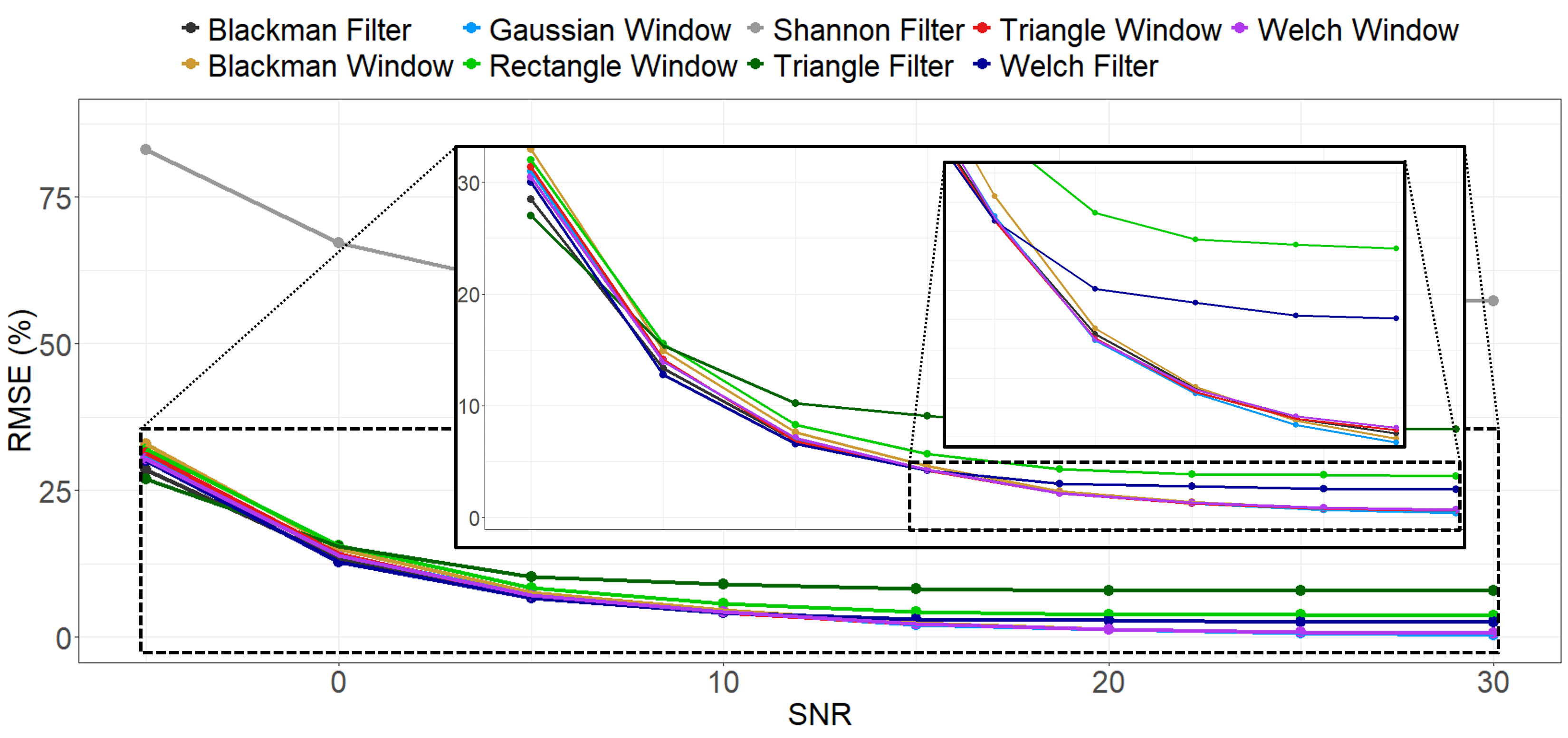}
  \caption{Scenario 1—\textbf{Top:} distribution of the estimated damping ratio for Mode~2 across SNR; the dashed line marks the true \(\zeta_2=1\%\). \textbf{Bottom:} RMSE(\%) versus SNR for all methods.}
  \label{fig:scenario1}
\end{figure}

Figure~\ref{fig:scenario1} summarizes the results for Scenario 1. The Shannon filter is a clear outlier at all SNRs: it consistently underestimates the damping ratio. Among the rest, Gaussian, Blackman, Triangle, and Welch windows, together with the Blackman filter, constitute the top tier across SNRs. At high SNRs, this group achieves the lowest RMSE and the tightest spreads. Most of the filter variants do not generally confer benefits here: the Welch filter is markedly worse than its window counterpart, and the Triangle filter offers no consistent advantage over the top tier group. At mid SNRs, the Blackman window begins to fall behind, whereas at low SNRs, the Blackman filter shows the highest noise robustness compared with the Gaussian, Triangle, and Welch windows. Although the Welch filter has the lowest RMSE at -5 SNR, it fails to estimate the damping ratio accurately at higher SNR levels.

\begin{figure}[H]
  \centering
  \includegraphics[width=\linewidth]{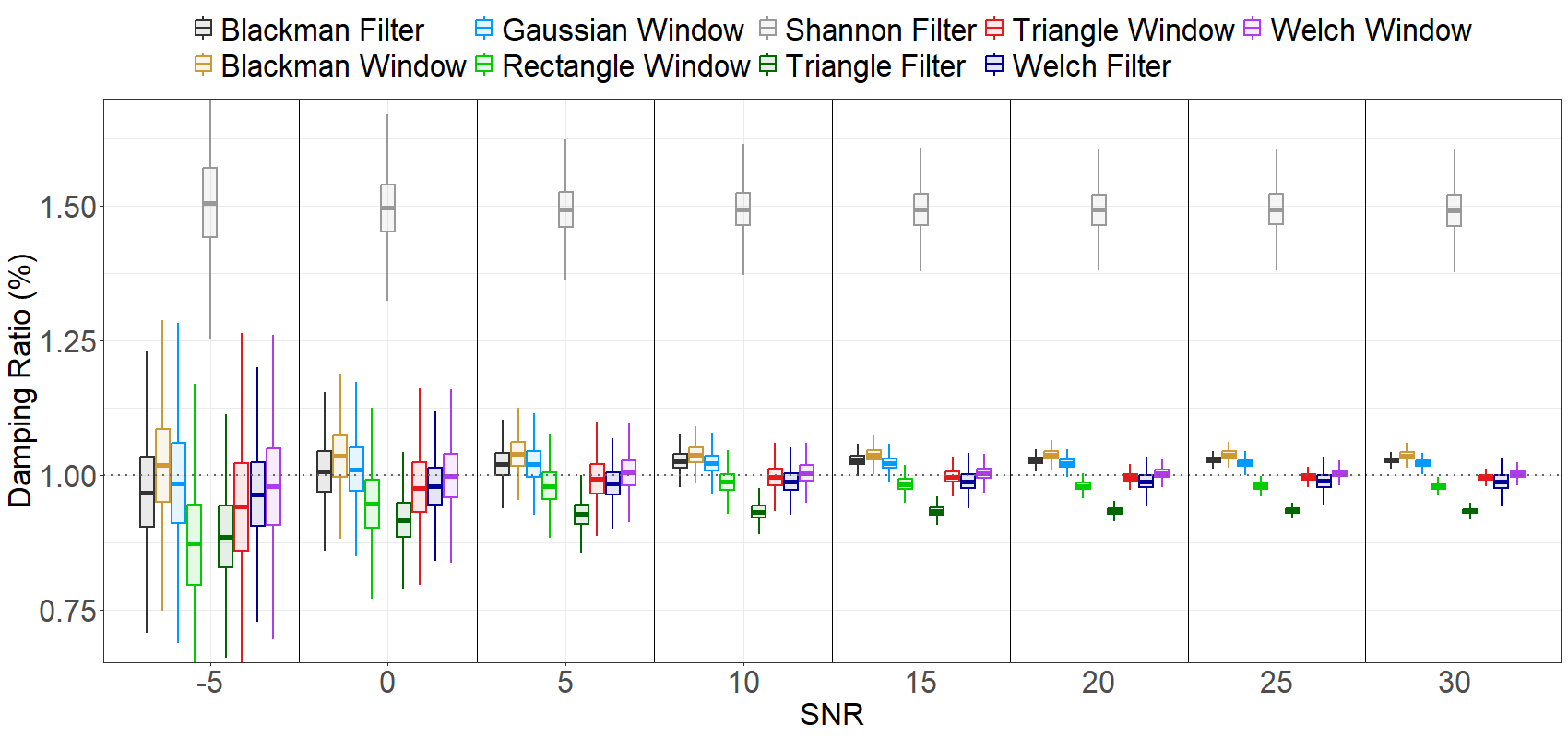}

  \medskip

  \includegraphics[width=\linewidth]{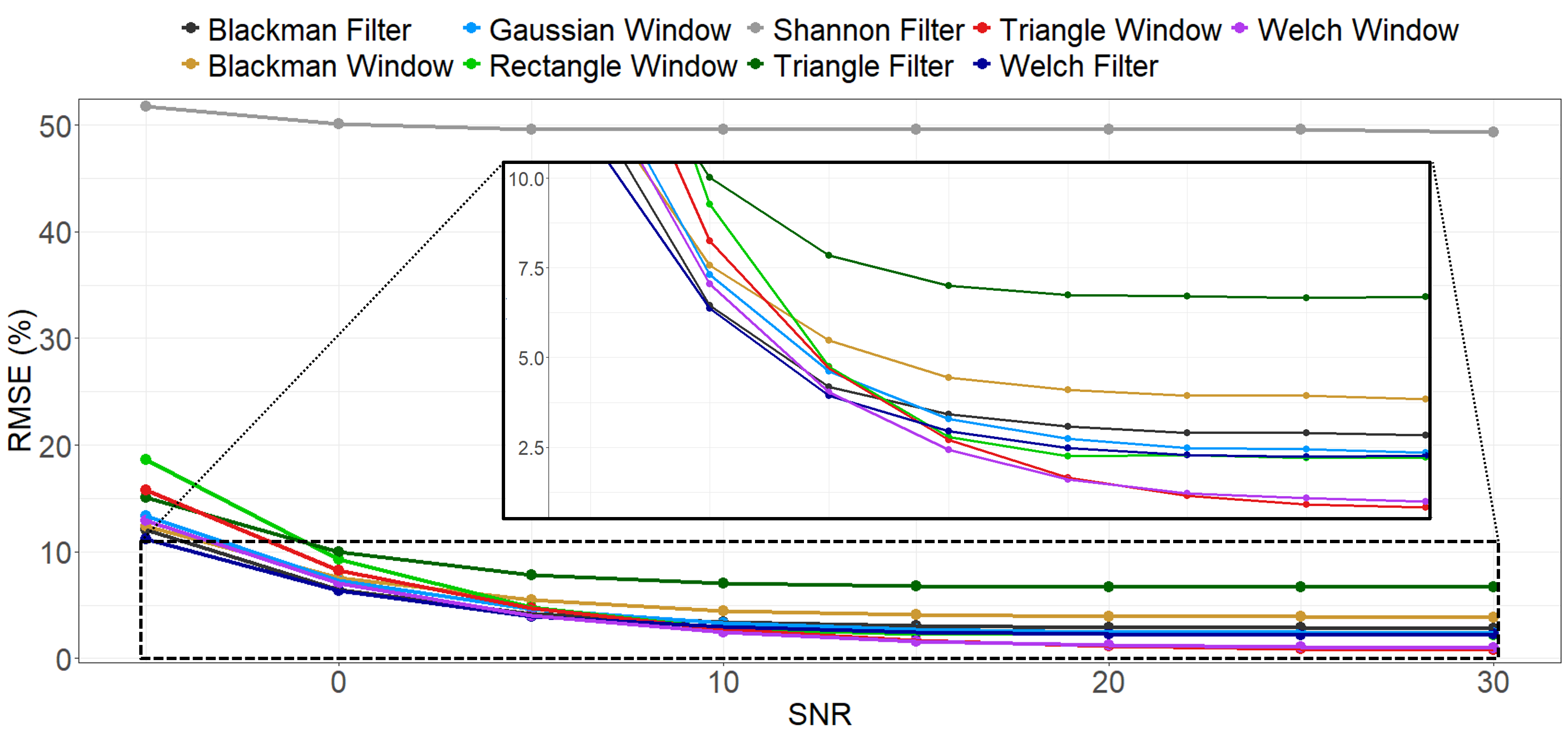}
  \caption{Scenario~2—\textbf{Top:} distribution of the estimated damping ratio (\%) for Mode~2 across SNR; the dashed line marks the true \(\zeta_2=1\%\). \textbf{Bottom:} RMSE(\%) versus SNR for all methods.}
  \label{fig:scenario2}
\end{figure}

For Scenario~2, the results are summarized in Fig.~\ref{fig:scenario2}. Starting from the high-SNR regime, the Triangle window exhibits the lowest errors, with the Welch window close behind; the Gaussian window, the Welch filter, and the Rectangle window trail this group slightly, followed by the Blackman filter.
At mid SNR, the Welch window starts surpassing the Triangle window, as the Triangle window starts degrading.
At mid SNRs, the Welch window surpasses the Triangle window as the latter begins to degrade.
In the low-SNR regime, the Welch and Blackman filters are comparatively robust, achieving the lowest RMSE; the Gaussian and Welch windows remain strong but degrade less than the Rectangle and Triangle windows. Shannon consistently maintains the highest error across all SNRs.
Overall, relative to the Gaussian window, the Welch window and the Welch filter consistently perform better across SNR levels. The Blackman filter often matches or exceeds the Gaussian (especially at very low SNRs), and the Triangle window is competitive at very high SNR. The Triangle filter shows systematic underestimation and higher error, and the Shannon filter is unsuitable under these conditions.

Scenario~3 represents a higher damping-ratio case (\(\zeta_2 = 4\%\)), which is inherently more challenging for the tested methods. Larger \(\zeta\) values produce faster exponential decay in the impulse response, resulting in shorter and lower-amplitude oscillatory tails. This reduces the number of usable cycles for slope-based envelope estimation and makes the extracted envelopes more sensitive to distortion from noise and windowing effects. Consequently, even at high SNRs, estimation variance and bias tend to be larger than in the lightly damped cases.

\begin{figure}[H]
  \centering
  \includegraphics[width=\linewidth]{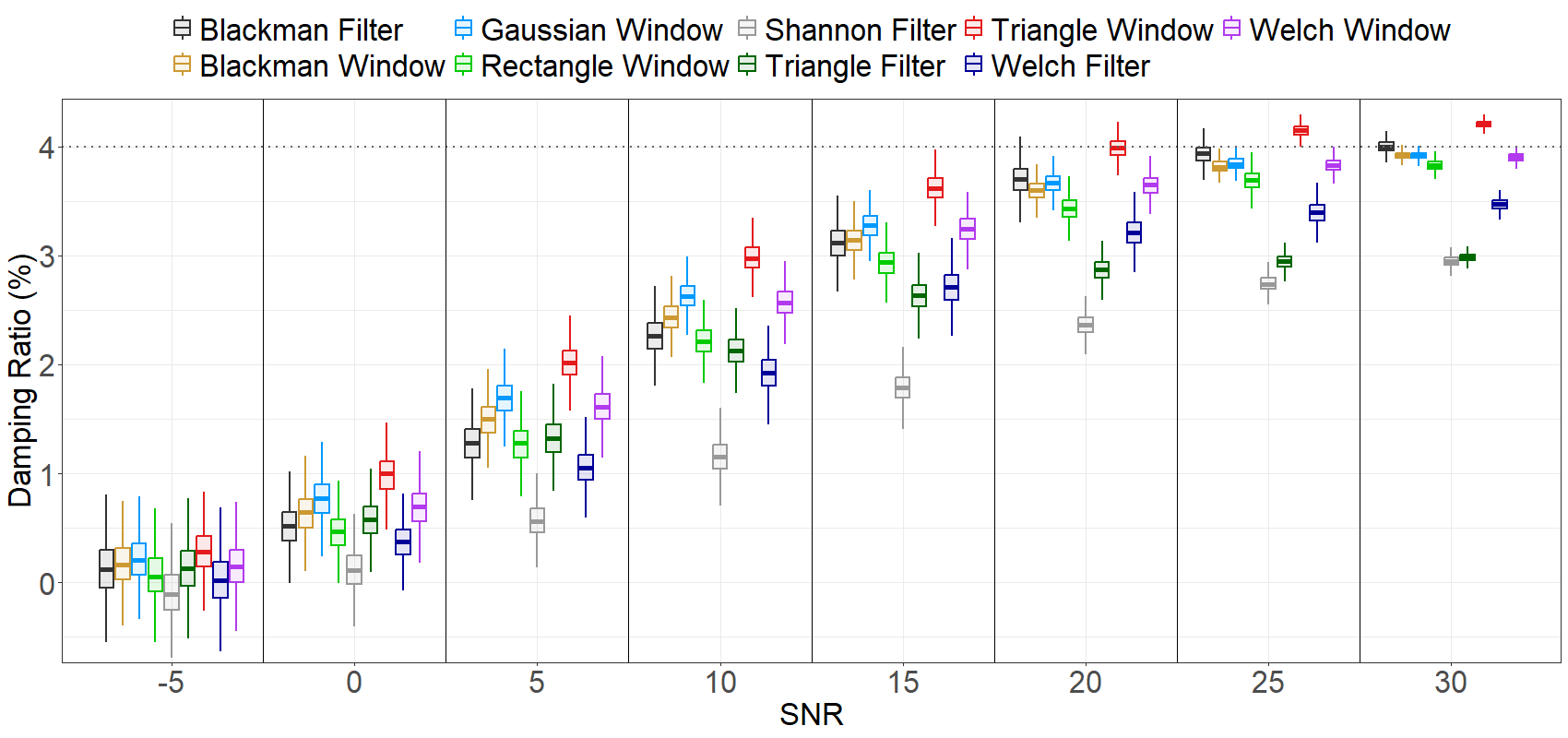}

  \medskip

  \includegraphics[width=\linewidth]{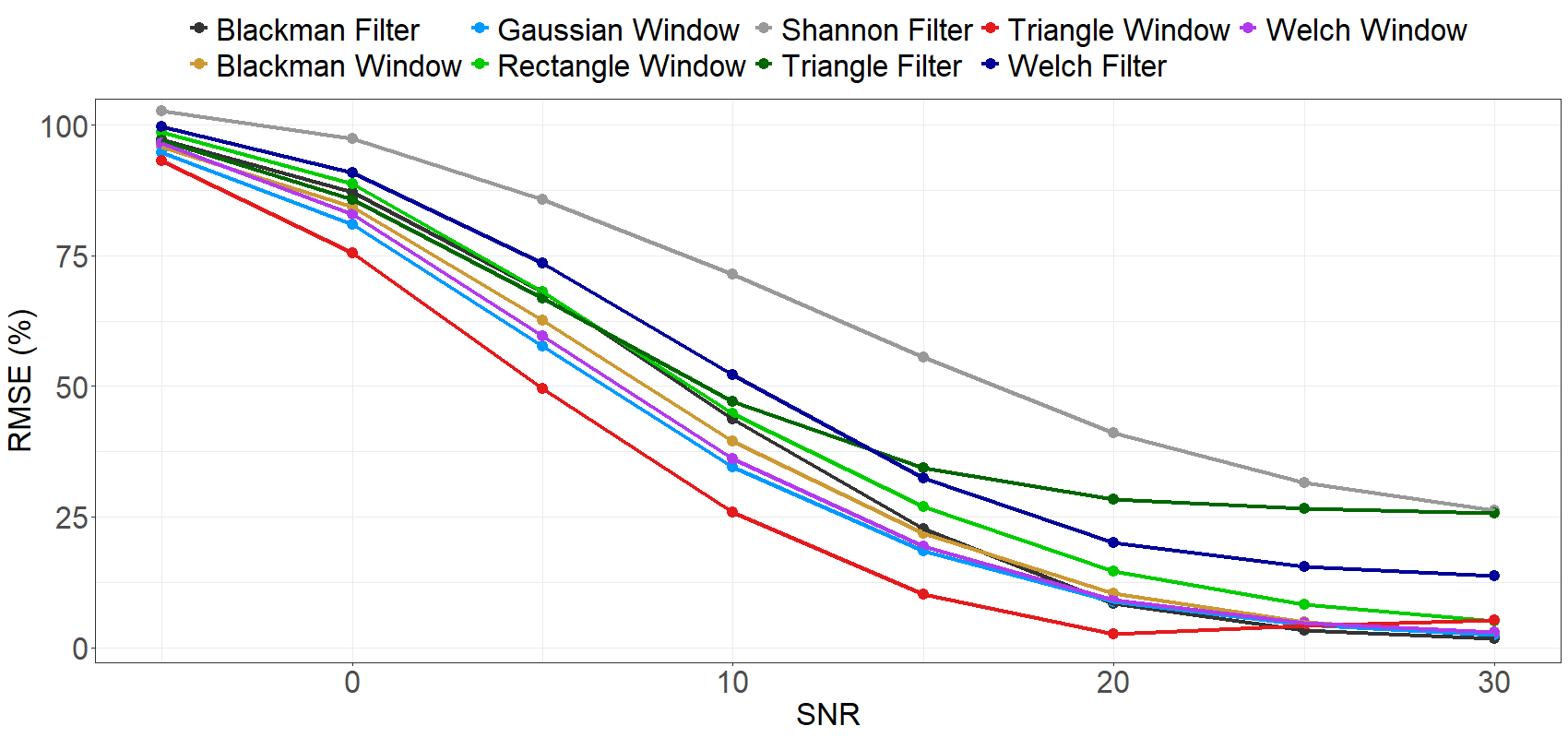}
  \caption{Scenario~3 (high damping)—\textbf{Top:} distribution of the estimated damping ratio (\%) for Mode~2 across SNR; the dashed line marks the true \(\zeta_2=4\%\). \textbf{Bottom:} RMSE(\%) versus SNR for all methods. The larger damping ratio makes accurate estimation more difficult, particularly at low SNR.}
  \label{fig:scenario3}
\end{figure}

Starting from the high-SNR end in Fig.~\ref{fig:scenario3}, the Blackman filter achieves the lowest RMSE, followed closely by the Gaussian, Blackman, Rectangle, and Welch windows. Most methods converge toward the true \(\zeta_2=4\%\) with reduced spread, although the Rectangle and Triangle windows remain slightly less accurate, and the Shannon filter continues to underestimate. The Welch and Triangle filters also show systematic negative bias, even at these SNRs.
At mid SNRs, the Triangle window retains a clear advantage in both the boxplots and RMSE curves. Gaussian, Welch, and Blackman windows remain competitive, while the Blackman filter begins to degrade significantly.
In the low-SNR regime, all methods exhibit large bias and variance, with medians far below the true value and RMSEs exceeding \(75\%\). The Triangle window still ranks best, with Gaussian, Welch, and Blackman windows performing better than the rest.
Overall, higher damping ratios amplify the challenge for envelope-based estimation by reducing the effective data length and robustness to noise. In this scenario, the Triangle window provides the best performance, especially across mid-to-high SNRs, with the Blackman filter and the Gaussian, Welch, and Blackman windows as strong alternatives at high SNRs.

\begin{figure}[H]
  \centering
  \includegraphics[width=\linewidth]{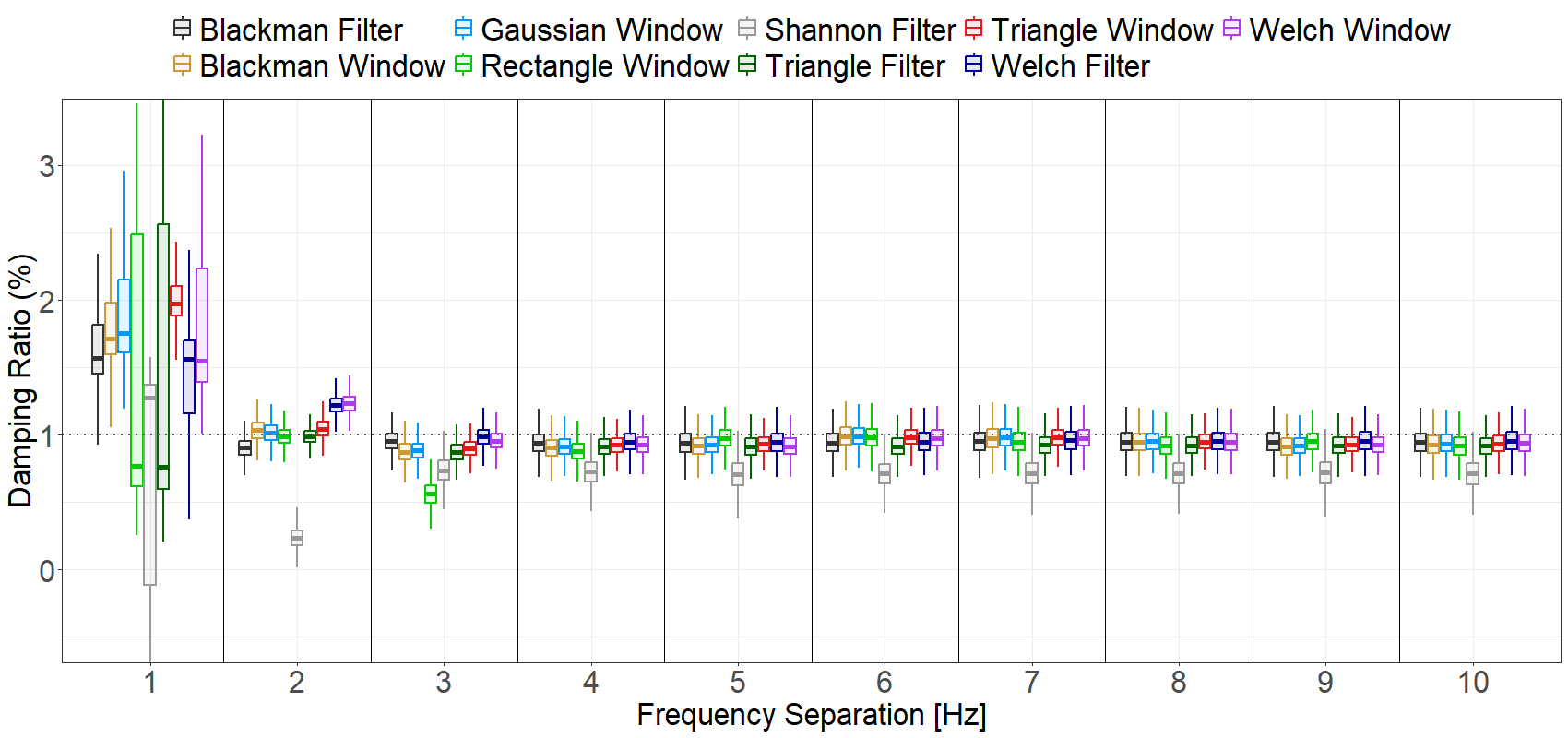}

  \medskip

  \includegraphics[width=\linewidth]{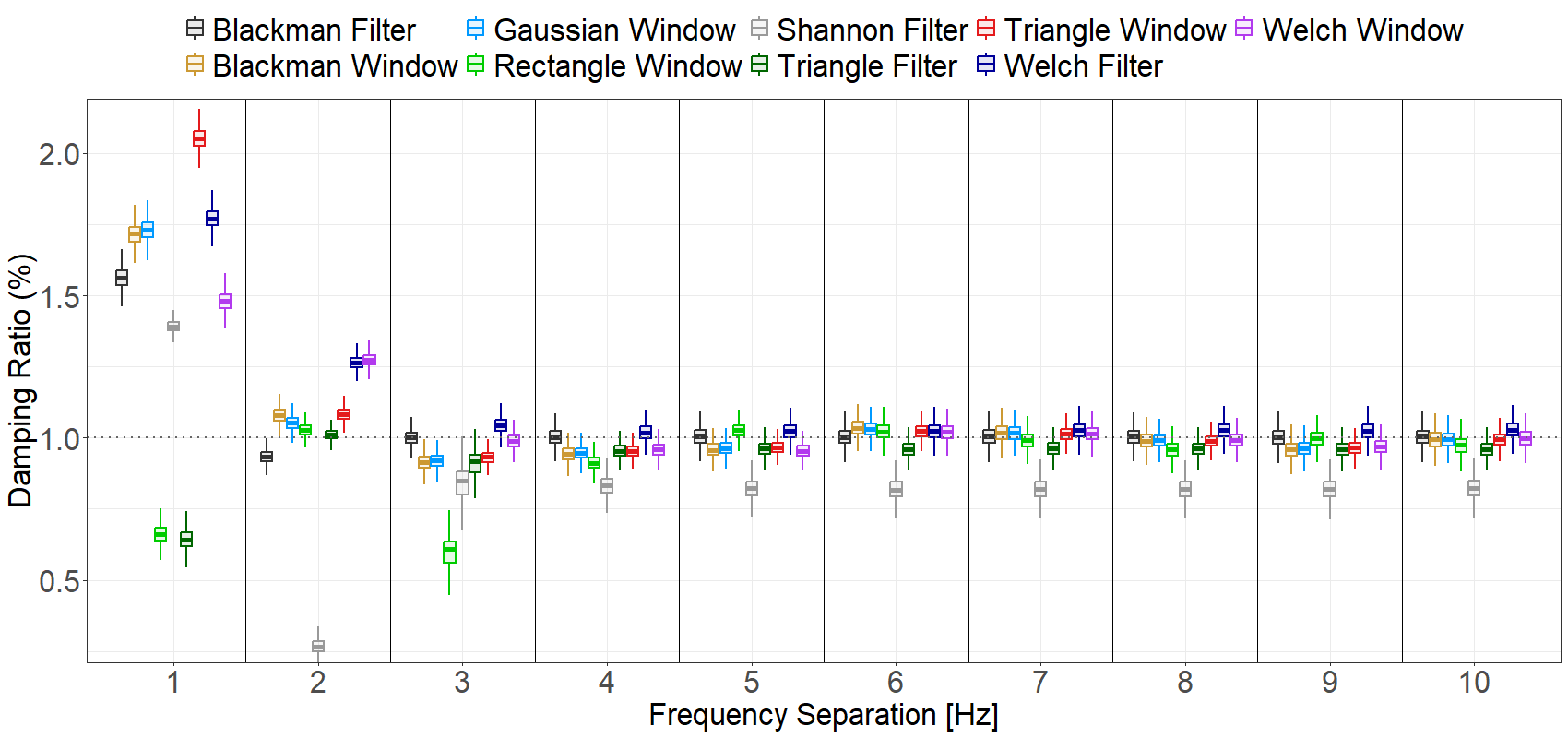}
  \caption{Effect of closely spaced modes—\textbf{Top:} SNR\,=\,0\,dB; \textbf{Bottom:} SNR\,=\,10\,dB. Each plot shows the estimated damping ratio (\%) for the target mode versus frequency separation \(\Delta f\in[1,10]\,\mathrm{Hz}\). The dashed line marks the true \(\zeta_2=1\%\).}
  \label{fig:close-modes}
\end{figure}

In the next experiment, we examined how closely spaced modes affect the performance of envelope estimators. To investigate mode interference, we fixed the target mode at \(\omega_2=15.56\,\mathrm{Hz}\) with a damping ratio of \(\zeta_2=1\%\) (as in Scenarios 1 and 3) and a unit modal amplitude. We then introduced two interfering modes, each with a damping ratio of \(\zeta_{1,3}=4\%\) and a modal amplitude of 5. The frequency separation \(\Delta f\) between each interferer and the target mode was varied from \(10\) to \(1\,\mathrm{Hz}\). For each method, we used parameters \(\theta\) that were optimized for scenarios 1 and 3 and evaluated the performance at two representative noise levels: 0 and 10 dB. Figure~\ref{fig:close-modes} illustrates the estimated \(\zeta_2\) as a function of \(\Delta f\) (the dashed line indicates the true value at \(1\%\)).

For \(\Delta f \gtrsim 6\,\mathrm{Hz}\), all methods show minimal effect, indicating negligible. Degradation begins around \(5\,\mathrm{Hz}\) for nearly all methods. Although the distributions begin to shift at \(5\,\mathrm{Hz}\), accuracy is not markedly affected until \(3\,\mathrm{Hz}\). Although some estimators, particularly the Blackman and Gaussian windows, remain usable at a \(2\,\mathrm{Hz}\) frequency separation, the Blackman filter is notably resilient in this close-mode setting; it shows no adverse effects at \(3\,\mathrm{Hz}\)and maintains good accuracy at \(2\,\mathrm{Hz}\).

\begin{figure}[H]
  \centering
  \includegraphics[width=\linewidth]{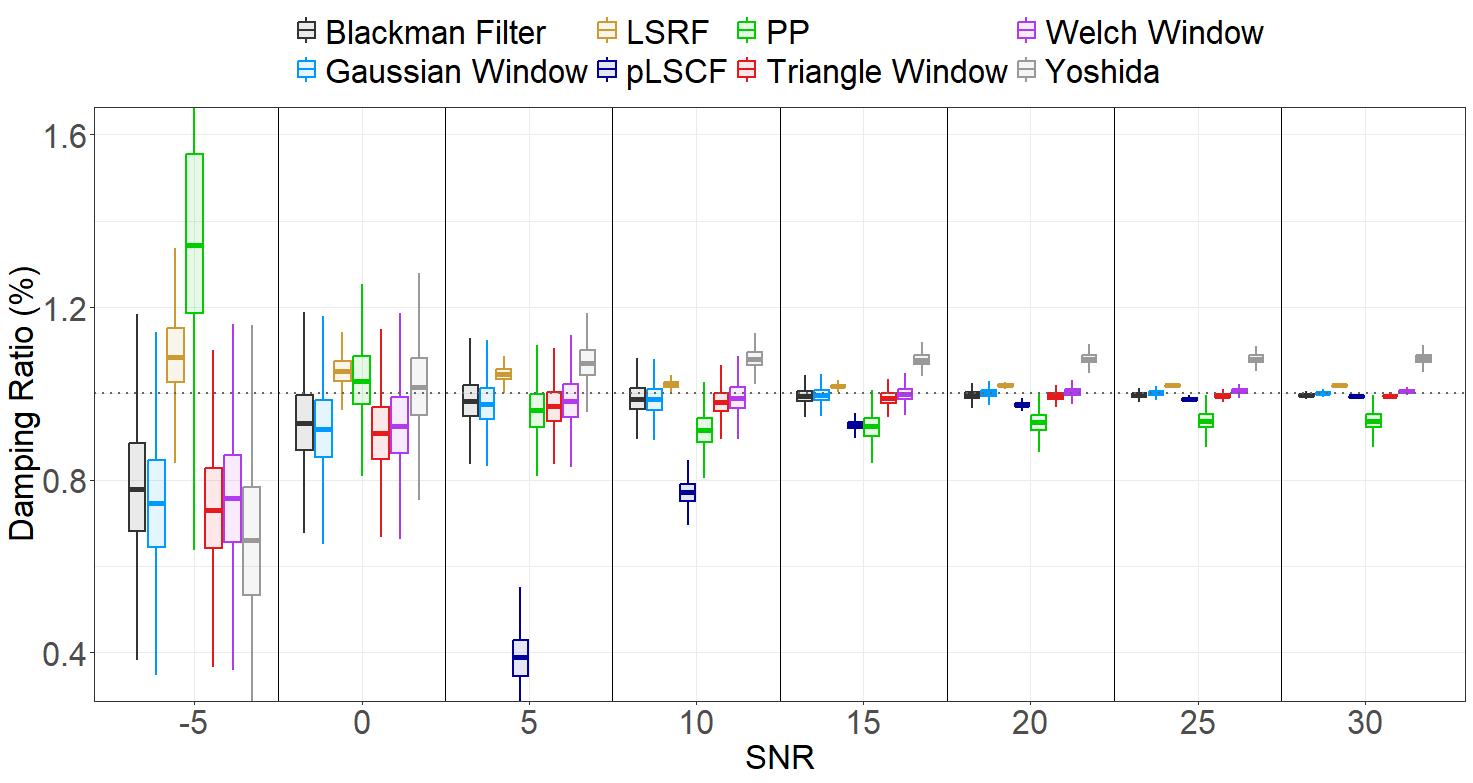}

  \medskip

  \includegraphics[width=\linewidth]{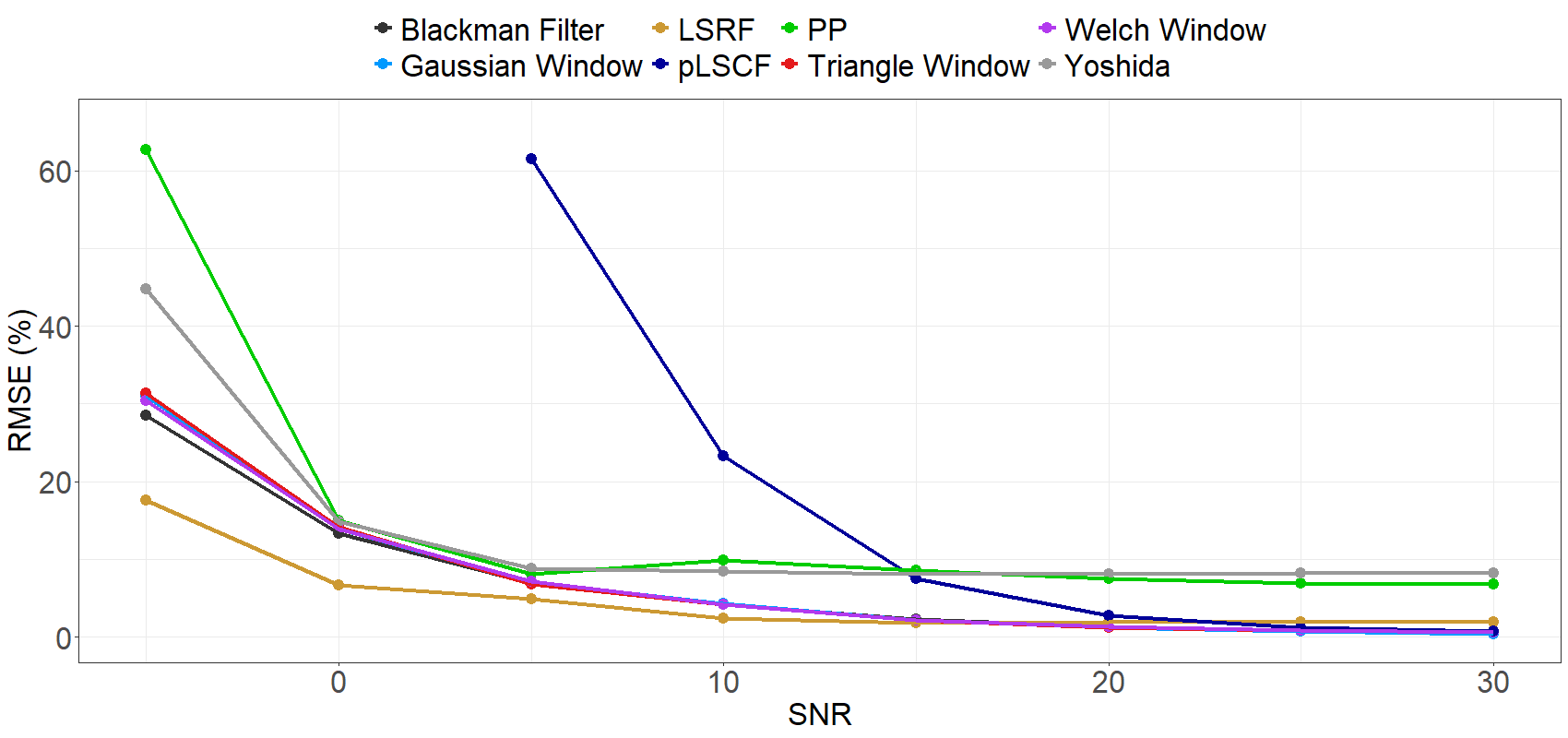}
  \caption{Scenario 1—Top: distribution of the estimated damping ratio for Mode 2 versus SNR; the dashed line marks the true \(\zeta_2=1\%\). Bottom: RMSE(\%) versus SNR comparing envelope estimators with LSRF, pLSCF, PP, and Yoshida.}
  \label{fig:fmethods-s1}
\end{figure}

We next compared the best envelope estimators identified above (Gaussian, Triangle, and Welch windows, plus the Blackman filter) with established frequency-domain baselines, LSRF, pLSCF, PP, and the Bertocco–Yoshida method, on the same Scenario 1 signals. Figure~\ref{fig:fmethods-s1} reports the boxplots and RSME.

As mentioned, the raw recordings in Dataset 1 are not time-aligned (each realization has an unknown impact time \(\tau_0\)). To provide a common reference, we estimated \(\tau_0\) as the time at which the envelope extracted by the Gaussian-window estimator attains its maximum. This single \(\hat{\tau}_0\) was then used for all methods in each realization to ensure a fair comparison.

With this estimate in hand, we tried two standard ways to construct the FRF for the frequency-domain methods.

\begin{enumerate}
    \item Discard all samples preceding the estimated impact time, compute the fast Fourier transform (FFT) of the remaining signal, and apply the chosen method directly to that spectrum. 
    \item Create a synthetic input that is a single impulse at the estimated impact time, take FFTs of both input and output, form their ratio to obtain the FRF, and then apply the method to that response.
\end{enumerate}
In our experiments, PP worked best with the first method, while LSRF, pLSCF, and Yoshida performed better with the second technique. We adopted these settings consistently when producing the results in Fig.~\ref{fig:fmethods-s1}.

Starting in the high-SNR regime, the envelope estimators converge near the true damping with very small spread and RMSE on the order of one percentage point. The LSRF shows a slight positive bias but remains competitive. pLSCF catches up only in this regime after being unstable at lower SNRs. Yoshida and PP retain noticeable bias and higher RMSE than the leading methods.
At mid-SNR, the envelope estimators deliver tight, near-unbiased estimates. However, LSRF remains extremely strong and typically has the lowest RMSE around 5–10 dB. pLSCF is notably unstable at 5–10 dB (severe underestimation and an RMSE spike). Yoshida and PP continue to underperform in this range due to bias.
At very low SNRs, LSRF is the most reliable option, yielding the smallest errors among all methods. The envelope estimators remain usable but exhibit broader spreads relative to higher SNRs. pLSCF is unstable in this regime, and its results were omitted from the charts. PP and Yoshida deviate markedly from the true value, while exhibiting large bias among the baselines.

Across the three scenarios and the closely spaced-mode and baseline comparisons, a consistent picture emerges. First, there is no universal advantage for Gaussian‐windowed wavelets (Morlet/Gabor) for envelope-based damping estimation. Triangle and Welch windows repeatedly match or exceed the Gaussian in both median accuracy and spread, particularly at moderate to high SNR. Also, the Blackman filter is often competitive and, in some low-SNR cases, more resilient than windowed envelopes. By contrast, the Shannon filter is systematically biased and is not recommended. Additionally, for closely spaced modes, interference increases as the separation decreases; among the tested methods, the Blackman filter shows the highest resistance to such interference.
Finally, against frequency-domain baselines, LSRF is the most reliable at very poor SNRs, but these envelope estimators match or surpass the frequency-domain methods from moderate SNR upward, whereas pLSCF is dependable only at high SNR, and PP/Yoshida tend to retain bias.

\section{Conclusion} \label{conclusion}

This work re-examined a common assumption in modal analysis: that Gaussian‐windowed wavelets (Morlet/Gabor) are the default choice for envelope extraction and damping estimation. Using a data-driven optimization framework and a systematic evaluation across SNR and modal spacing, benchmarking against frequency-domain baselines, we found no universal advantage of Gaussian-windowed wavelets in envelope-based damping estimation. In many practically relevant conditions, alternatives such as the Triangle and Welch windows, and in some cases the Blackman filter, achieve equal or better accuracy. 

From a practitioner’s perspective, method selection should follow operating conditions rather than convention: when SNR is very low in lightly damped systems, LSRF is the most robust; as SNR improves, envelope-based estimators with Triangle or Welch windows (and, in some cases, a Blackman filter) become preferable; when modes are closely spaced, the Blackman filter shows the highest resilience to mode interference; and under strong damping, envelope estimators generally degrade due to shortened decays, but the Triangle window stands out as the most reliable among them.

The study has some limitations. The experiments were conducted using LTI systems and synthetic responses with controlled noise and alignment procedures. Real-world systems, however, may exhibit nonlinearities, nonstationarity, imperfect excitation, and other challenges. Our focus was on estimating mode-specific damping. Future work will extend these findings to nonlinear systems and real experimental datasets.

Overall, the evidence indicates that Gaussian-windowed wavelets (Morlet/Gabor) should not be treated as the default for damping estimation via envelopes. Alternatives, particularly Triangle and Welch windows and, in some regimes, the Blackman filter, often yield lower errors and greater robustness.


\bibliographystyle{elsarticle-num}
\bibliography{sample}

\end{document}